\documentclass[twocolumn,nofootinbib,amsmath,amssymb,aps,prd,balancelastpage,superscriptaddress]{revtex4-1}

\usepackage[english]{babel}
\usepackage[utf8]{inputenc}
\usepackage[colorinlistoftodos, color=green!40, prependcaption]{todonotes}
\usepackage{amsmath}
\usepackage{amsfonts}
\usepackage{amssymb}
\usepackage{dsfont}
\usepackage{eqnarray,amsmath}
\makeatletter\AtBeginDocument{\let\@elt\relax}\makeatother
\usepackage{graphicx}
\usepackage{bbold}

\usepackage[colorlinks=true,linkcolor=blue,urlcolor=blue,citecolor=blue]{hyperref}
\usepackage[normalem]{ulem}

\begin{document}

\title{Cancellation effects as a fingerprint of quantum collapse models at atomic scale}

\author{Kristian Piscicchia}
\affiliation{Centro Ricerche Enrico Fermi -- Museo Storico della Fisica e Centro Studi e Ricerche ``Enrico Fermi'', Italy} 
\affiliation{Laboratori Nazionali di Frascati, INFN, Italy;}
\author{Sandro Donadi}
\affiliation{Centre for Quantum Materials and Technologies, School of Mathematics and Physics,
Queen’s University, Belfast BT7 1NN, United Kingdom;}
\author{Simone Manti}
\affiliation{Laboratori Nazionali di Frascati, INFN, Italy;}
\author{Angelo Bassi}
\affiliation{Department of Physics, University of Trieste, Italy}
\affiliation{Istituto Nazionale di Fisica Nucleare,~Section of Trieste, Italy;}
\author{Maaneli Derakhshani}
\affiliation{Department of Mathematics, Rutgers University -- New Brunswick, New Jersey, USA;}
\author{Lajos Diósi}
\address{E\"{o}tv\"{o}s Lor\'and University, Budapest, Hungary}
\address{Wigner Research Centre for Physics, Budapest, Hungary}
\author{Catalina Curceanu}
\affiliation{Laboratori Nazionali di Frascati, INFN, Italy;}
\affiliation{IFIN-HH, Institutul National pentru Fizica si Inginerie Nucleara Horia Hulubei, Romania.}
\begin{abstract}
\noindent


In this work the spontaneous electromagnetic radiation from atomic systems, induced by dynamical wave-function collapse, is investigated in the X-rays domain. Strong departures are evidenced with respect to the simple cases considered until now in the literature, in which the emission is either perfectly coherent (protons in the same nuclei) or incoherent (electrons). In this low-energy regime the spontaneous radiation rate strongly depends on the atomic species under investigation and, for the first time, is found to depend on the specific collapse model.

\end{abstract}

\maketitle

{\it Introduction.} ---

Quantum mechanics is our most successful physical theory, allowing us to understand and predict a large number of phenomena with extreme precision \cite{fan2023measurement}. At the core of the theory lies the superposition principle, according to which systems are allowed to be in superposition of different states. The thoery does not set any boundary on the limit of validity of this principle; yet, we don't observe superpositions of macroscopic objects. This well-known problem is exemplified by the Sch\"ordinger's cat Gedankenexperiment \cite{schrodinger1935gegenwartige}.\\

It has been suggested that the linearity of quantum mechanics, from which the superposition principle directly follows, may break down at a certain scale \cite{leggett1980macroscopic,weinberg1989precision,bell2004speakable}. This idea has been extensively developed in the framework of collapse models \cite{pearle1976reduction,ghirardi1986unified,bassi2003dynamical,bassi2013models}, which are phenomenological models that modify the Schr\"odinger equation by adding non-linear and stochastic terms that naturally collapse the wave function in space. According to these models, microscopic systems are very weakly affected by the non-linearities which, however, become dominant when atoms glue together to form larger and larger systems, this way solving the measurement problem.
Among the several collapse models proposed, two are of particular relevance: the Continuous Spontaneous Localization (CSL) model \cite{ghirardi1990continuous} and the so-called Di\'osi-Penrose (DP) model \cite{diosi1989models,penrose1996}.


Spontaneous collapses must be random, in order to avoid the possibility of faster than light signaling \cite{gisin1989stochastic}; this randomness manifests as a diffusive motion of the system \cite{donadi2023collapse}, which corresponds to a random acceleration of the atoms and, hence, to the emission of radiation from their charged constituents.
The experimental search of this spontaneous radiation was performed, for both CSL \cite{donadi2021novel} and DP \cite{donadi2021underground}, in the energy domain of the $\gamma$-rays, by comparing the measured radiation spectrum by high purity germanium crystals with the spontaneous emission rate predicted by the models for the atomic systems which constitute the experimental setup. The obtained strong bounds, combined with constraints provided by other experimental tests and theoretical considerations, 
are leading to a progressive falsification of the models in their Markovian formulation.\\

Theoretical efforts have being devoted to the development of non-Markovian collapse models \cite{adler2007photon, bassi2009electromagnetic, adler2013spontaneous,bassi2014spontaneous,donadi2014spontaneous}, in order to counteract the runaway energy increase. These new models require the introduction of a cutoff frequency in the stochastic noise spectrum; for this reason, a systematic scan of the spontaneous radiation phenomenon, as a function of the decreasing energy, is mandatory.

The search for spontaneous radiation emitted by Germanium crystals was performed in the X-rays domain in \cite{piscicchia2017csl} (for $E\in (15-50)$ keV), and more recently in \cite{majorana} (for $E\in (19-100)$ keV). In \cite{piscicchia2017csl} a formula for the expected
spontaneous radiation rate from quasi-free electrons was applied: the expected radiation depends on the
 energy as $1/E$ and is proportional to the number of quasi-free electrons \cite{fu}. 
This formula is not suitable to describe the more complex phenomenology of the spontaneous radiation
emitted by the whole atomic system;
more refined calculations are presented in \cite{donadi2021novel,donadi2021underground}, where the CSL and the DP rates are calculated for an atomic system, in the limit in which the spontaneous photon wave-length $\lambda_{\gamma}$ is between the nuclear dimension and the mean radius of the lower laying atomic orbit. Consistently, in \cite{donadi2021novel,donadi2021underground} the data analyses are performed for photons energies in the range (1-3.8) MeV.  The rate results to be proportional, for both CSL and DP, to $(N_p^2 + N_e)/E$, where $N_p$ and $N_e$ are, respectively, the number of protons and electrons of the atom under study. In this regime, different collapse models share the same expected shape for the energy distribution of the spontaneous emission rate, the scaling factor being proportional to combinations of constants of nature with the characteristic parameters of the models, which are $\lambda$ and $r_C$ (strength and correlation length of the collapse noise) for CSL and the correlation length $R_0$ for DP (the role of the strength being played by the gravitational constant $G$ in this case).   
The latter theoretical rates were also applied in \cite{majorana}, where the X-rays spectrum measured by Germanium crystals is analyzed in the energy range (19-100) keV, in which $\lambda_{\gamma}$ is comparable with the mean radii of the atomic orbits of Germanium.

Given the importance of integrating the search of the spontaneous radiation signal in the high-energy domain of the $\gamma$-rays to the one in the X-Rays, we derive in this work the general expression of the radiation emission rates, for both Markovian and non-Markovian formulations of the CSL and DP models. We adopt a semiclassical approach which is valid above 1 keV (see e.g. \cite{adler2007photon}), appropriate to the current experimental surveys; for lower energies a fully quantum mechanical analysis is required, which is under development.
The general rate is found to exhibit a non-trivial energy dependence, which is strongly influenced by the interplay between the photons wavelengths, the radii of the electronic orbits and the correlation length of the model under study.   

Interestingly, the spontaneous radiation energy spectrum is found, at the atomic $\lambda_\gamma$ scale, to depend on the specific model of wave function collapse under scrutiny. This finding opens new scenarios in the experimental investigation of the spontaneous radiation: a measurement sensitive to this signature of the collapse, would be able to recognize the most probable pattern of dynamical wave function reduction.

{\it CSL spontaneous emission rate, general expression.} --- The rate of the spontaneous radiation emitted by an atomic system, in the context of the Markovian CSL model, was derived in \cite{donadi2021novel}:
\begin{equation}\label{ratecsl1}
\left. \frac{d\Gamma}{dE} \right|_t^{CSL} =   \frac{\hbar \, \lambda}{6\, \pi^2 \, \epsilon_0 \, c^3 \, m_0^2 \, E} \sum_{i,j} \frac{q_i \, q_j}{m_i \, m_j} \cdot f_{ij} \cdot \frac{\sin(b_{ij})}{b_{ij}},
\end{equation}
where $b_{ij}=2\pi |\mathbf{r}_{i}-\mathbf{r}_{j}| / \lambda_\gamma$, $q_{j}$ and $m_{j}$ represent, respectively, the charge and the mass of the $j$-th particle, at position $\mathbf{r}_{j}$. $m_0$ denotes the nucleon mass, 
$\epsilon_0$ the vacuum permittivity, $\hbar$ and $c$ are, as usual, the reduced Planck constant and the speed of light, $E$ is the energy of the spontaneously emitted photon. 

The term $f_{ij}$ encodes the balance between the emitters' distances and the correlation length $r_C$. A generalized expression for $f_{ij}$ is provided in  Appendix \ref{fij}:

\begin{equation}\label{fijcsl2}
f_{ij} = \frac{m_i \, m_j}{2 r_C^2} \, e^{-\frac{(\mathbf{r}_i-\mathbf{r}_j)^2}{4r_C^2}} \left( 3 - \frac{(\mathbf{r}_i-\mathbf{r}_j)^2}{2r_C^2} \right).
\end{equation}
Using Eq. \eqref{fijcsl2}, and analyzing the contributions to the spontaneous radiation by the protons in the nucleus, the orbital electrons and due the combined electrons-protons emission - see Appendix \ref{whiterate} - Eq. \eqref{ratecsl1} turns to:
\begin{widetext}
\begin{align}\label{ratecsl5}
&\left. \frac{d\Gamma}{dE} \right|_t^{CSL} \!\!=  \frac{\hbar \, e^2 \, \lambda}{12\, \pi^2 \, \epsilon_0 \, c^3 \, m_0^2\, r_C^2 \, E} \left\{
3\,N_p^2 + 3\,N_e + 2 \sum_{o \, o' \, \textrm{pairs}} N_{o} \, N_{o'} \, \frac{\textrm{sin} \left[ \frac{\beta|\rho_o-\rho_{o'}| \, E}{\hbar \, c} \right]}{\left[ \frac{\beta|\rho_o-\rho_{o'}| \, E}{\hbar \, c} \right]} e^{-\frac{\beta^2(\rho_o-\rho_{o'})^2}{4r_C^2}} \, \left(  3 - \frac{\beta^2(\rho_o-\rho_{o'})^2}{2r_C^2}  \right)-\right.
\nonumber\\
&\left.  -2 N_p \sum_o N_{o} \, \frac{\textrm{sin} \left( \frac{\rho_o \, E}{\hbar \, c} \right)}{\left( \frac{\rho_o \, E}{\hbar \, c} \right)} \cdot 
 e^{-\frac{\rho_o^2}{4r_C^2}} \, \left(  3 - \frac{\rho_o^2}{2r_C^2}  \right)
 + \sum_o N_{o} \, (N_{o} -1) \, e^{-\frac{(\alpha \, \rho_o)^2}{4r_C^2}} \cdot \frac{\textrm{sin} \left( \frac{\alpha \, \rho_o \, E}{\hbar \, c} \right)}{\left( \frac{\alpha \, \rho_o \, E}{\hbar \, c} \right)} \cdot \left(  3 - \frac{( \alpha \, \rho_o)^2}{2 r_C^2}  \right)
 \right\},
\end{align}
\end{widetext}

where the electron pairs distances are parametrized in terms of the mean radii of the atomic orbits  $\rho_o$ ($\rho_{o'}$) by means of the constants $\alpha$ and $\beta$ which are given below. $N_o$ represents the number of electrons in the $o$-th orbit of the atom.
Eq. \eqref{ratecsl5} represents a generalization of the spontaneous emission rate which was derived in Ref. \cite{donadi2021novel}:
\begin{equation}\label{ratecsl3}
\left. \frac{d\Gamma}{dE} \right|_t^{CSL} =   \frac{\hbar \, e^2 \, \lambda}{4\, \pi^2 \, \epsilon_0 \, c^3 \, r_C^2 \, m_0^2 \, E} \left( N_p^2 + N_e \right),
\end{equation}
under the condition that $\lambda_{\gamma}$ is intermediate between the nuclear and atomic dimensions,
i.e. the energy range under scrutiny belongs to the $\gamma$-rays domain. In that case protons emit coherently (proportionally to the square of their number) and electrons emit independently (linear dependence). More complex is the situations described by Eq. \eqref{ratecsl5}.\\

If $r_C$ exceeds the distance between the emitters (as confirmed by radiation experiments for a Markovian CSL \cite{donadi2021novel,piscicchia2017csl}) then the stochastic field ``shakes" them coherently. If $\lambda_\gamma$ becomes also of the order of the mean orbit radii of the atom, then the electrons of the corresponding orbits start to emit coherently, i.e. quadratically. Nonetheless, the corresponding increase in the expected spontaneous emission rate is counteracted by the \emph{cancellation}, among oppositely charged particles whose distance is smaller than $\lambda_\gamma$.  In the limit in which $\lambda_\gamma$ is also much bigger than the atomic size,  Eq. \eqref{ratecsl5} reduces to:
\begin{equation}
\left. \frac{d\Gamma}{dE} \right|_t^{CSL} = 
 \frac{\hbar \, e^2 \, \lambda}{4\, \pi^2 \, \epsilon_0 \, c^3 \, m_0^2\, r_C^2 \, E} \left[
N_p^2 - 2 \cdot N_p \, N_e 
+
N_e^2,
\right]
\end{equation}
which vanishes for neutral atoms.

\begin{figure}
\centering
\includegraphics[width=\columnwidth]{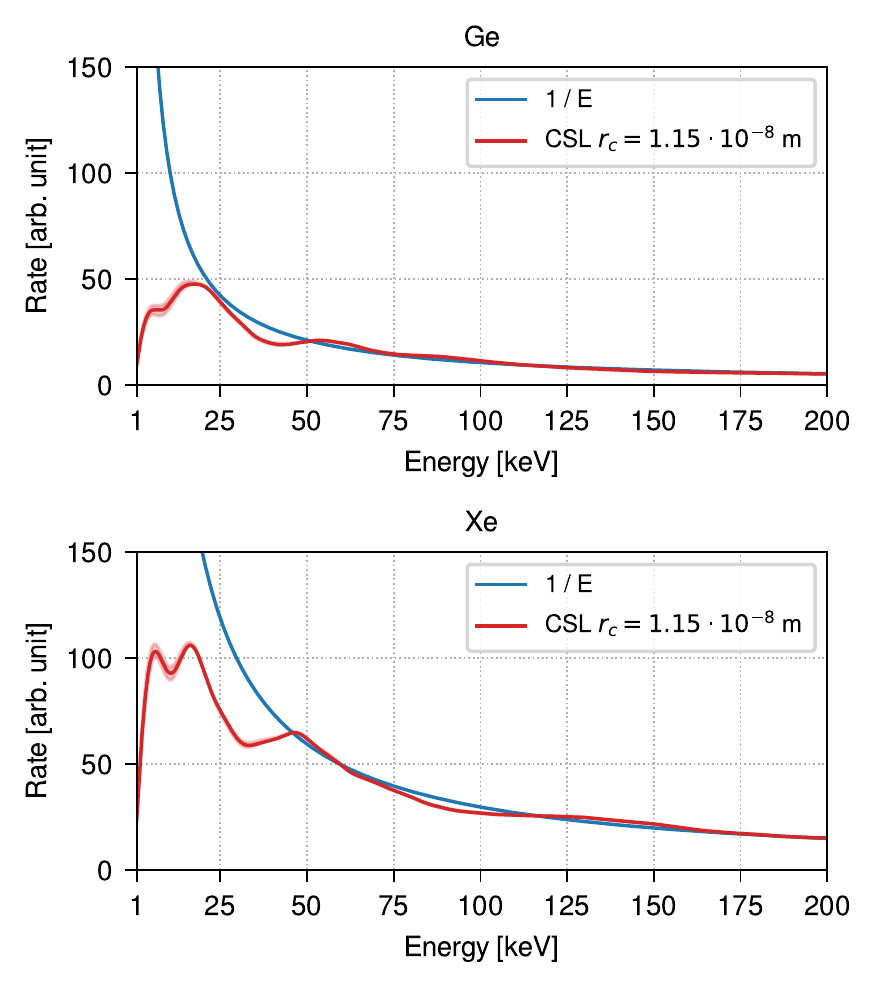}
\caption{The top panel of the Figure shows in blue the $1/E$ dependence Eq. \eqref{ratecsl3}, for the spontaneous radiation rate of a CSL model, which is only valid in the high-energy domain. This is compared to the general rate in Eq. \eqref{ratecsl5} (red curve) for a prior value of the correlation length $r_C = 1.15 \cdot 10^{-8}$ m. The distributions are calculated for a Germanium atom and normalized to the common constant pre-factors.
 The bottom panel of the Figure shows the shapes of the same rates, calculated for a Xenon atom.}
\label{comparison}
\end{figure}

Intermediate regimes for $\lambda_\gamma$ and $r_C$, in comparison with $|\mathbf{r}_i-\mathbf{r}_j|$, give rise to a new, interesting pattern, in which the shape of the expected spontaneous radiation spectrum exhibits a non-trivial energy dependence, which is influenced by the atomic structure.  
This is exemplified in Fig. \ref{comparison}. The top panel of Fig. \ref{comparison} shows in red the general spontaneous emission rate predicted by Eq. \eqref{ratecsl5} (red), compared to the simple (blue) case described by Eq. \eqref{ratecsl3}, for Germanium (Ge targets were indeed used in various experiments, e.g. \cite{donadi2021novel,majorana}). 
The red shaded area corresponds to the values  $\beta = 1.04$ and $\alpha$ spanning in the range (1 - 1.5), according to the literature \cite{gill2003,coulson}, see Appendix \ref{whiterate}.
The same rates for a Xenon target are shown in the bottom panel of Fig. \ref{comparison}.  In Eq \eqref{ratecsl5} the value of $r_C$ is set to $1.15 \cdot 10^{-8}$ m, consistently with the results of Ref. \cite{donadi2021novel}, which are obtained by applying Eq. \eqref{ratecsl3} in the $\gamma$-rays regime ($(1-3.8)$ MeV), where Eq. \eqref{ratecsl3} is an excellent approximation. $r_C=1.15\cdot 10^{-8}$ m corresponds to the intersection among the experimental bound and the theoretical constraint (corresponding respectively to the orange and gray lines in Fig. 4 of Ref. \cite{donadi2021novel}).  The mean radii of the orbits are obtained based on a density functional theory (DFT) \cite{kohmsham} all-electron calculation, for an isolated atom; the DFT code GPAW \cite{Enkovaara_2010} is adopted. 
The distributions are normalized to the common constant pre-factors to evidence differences in shape. 
As expected, the simple (blue) and the general (red) rates converge for high energies (above 200 keV), where $\lambda_\gamma$ becomes sizably smaller than the lower atomic orbit radii. Since $r_C$ is much greater than the size of the Germanium atom, the X-Rays regime is, instead, characterized by a balance among electrons and protons coherent emission and the cancellation of their contributions. The analysis performed in Ref. \cite{majorana} should be re-considered based on this low-energy complex pattern.   

The general expression of the spontaneous emission rate in Eq. \eqref{ratecsl5} encodes the phenomenology for future investigations of the spontaneous radiation at low energies (X-rays). Comparison of the theoretical expectation with the measured spectra requires a recursive analysis: in the first step, a suitable prior has to be assumed for $r_C$, an updated value for $r_C$ will be obtained, which will serve as input for the new prior. The analysis should then be iterated till convergence of the $r_C$ values, within the experimental sensitivity, is reached.

The dependence of the expected rate on the atomic structure becomes evident comparing the top panel of Fig. \ref{comparison} with the bottom one, which shows in red colour the general spontaneous emission rate given by Eq. \eqref{ratecsl5} for Xenon (high sensitivity bounds on the spontaneous collapse could be set by the XENON experiment \cite{aprile2019xenon1t} by exploiting a Xenon target); the blue line describes again the simple rate of Eq. \eqref{ratecsl3}. Eq. \eqref{ratecsl5} predicts a strong dependence of the spontaneous radiation yield on the atomic number $Z$; as such a survey of the spontaneous collapse induced emission over $Z$ would greatly improve the experimental sensitivity on this, new physics, phenomenon.

The generalization of Eq. \eqref{ratecsl5} to the non-Markovian case requires to multiply the right-hand side by the Fourier transform of the noise correlation function (see e.g. \cite{adler2007photon,adler2013spontaneous,bassi2014spontaneous,donadi2014spontaneous,carlesso2018} and the derivation of the colored DP model emission rate below). Assuming e.g. an exponentially decaying noise correlation function ($f(t-s) = \frac{\Omega}{2} \, e^{-\Omega|t-s|}$), characterized by a correlation time $\Omega^{-1}$, the rate becomes:
\begin{equation}\label{ratecslcol}
\left. \frac{d\Gamma}{dE} \right|_t^{cCSL} = \left. \frac{d\Gamma}{dE} \right|_t^{CSL}   \times \frac{E_c^2}{E_c^2 + E^2},
\end{equation}
where $E_c = \hbar \Omega$ and cCSL denote results for a colored (non-Markovian) CSL model.

{\it DP spontaneous emission rate, general expression.} ---
The rate of the spontaneous radiation emitted by an atomic system, according to a Markovian DP model, was derived in Ref. \cite{donadi2021underground}:
\begin{equation}\label{ratewdpnaive}
\left. \frac{d\Gamma}{dE} \right|_t^{DP} = \frac{Ge^2}{12 \pi^{5/2} \epsilon_0 c^3 R_0^3 E} \left( N_p^2 + N_e \right).
\end{equation}
assuming a spontaneous photon wavelength  
which is much bigger than the nuclear size and much smaller than the lower lying atomic orbit mean radius.
Note that Eq. (\ref{ratewdpnaive}) differs from the result presented in \cite{donadi2021underground} by a factor $8\pi$. This is because in \cite{donadi2021underground} we adopted the convention introduced in \cite{penrose1996} while we refer here to the original model introduced by Di\'osi \cite{diosi1987universal}, in which the factor $8\pi$ was not present.

The general structure of the rate is derived in Appendix \ref{ratedpc}, where the non-Markovianity of the noise time correlation is also considered. For an exponential time correlation we have:
\begin{eqnarray}\label{rate}
&&\left. \frac{d\Gamma}{dE} \right|_t^{cDP} = \frac{G}{6\pi^2\epsilon_{0}c^{3}E}  \sum_{i,j} q_i\, q_j  f_{ij} \frac{\textrm{sin}(b_{ij})}{b_{ij}} \frac{E_c^2}{E_c^2 + E^2}=
\nonumber\\
&& = \left. \frac{d\Gamma}{dE} \right|_t^{DP}  \frac{E_c^2}{E_c^2 + E^2},
\end{eqnarray}
with $G$ the Newton constant. 
The rate for the Markovian model is recovered in the limit $E_c \rightarrow \infty$. 
In analogy with the expected rate for the CSL model (Eq. \eqref{ratecsl1}), the interplay between the particles mean distance and the wavelength of the spontaneously emitted photon is contained in the terms $\textrm{sin}(b_{ij})/b_{ij}$. The dependence on the particles distances in relation to the correlation length of the model $R_0$ is instead specified by the terms $f_{ij}$.
As it is shown in Appendix \ref{ratedpc2}, $f_{ij}$ is a measure of the overlap between the mass densities of the particles $i$-th and $j$-th ($g_{i,j}$), whose spatial resolution is measured by $R_0$. In formulae:
\begin{align}\label{fij1}
    f_{ij}=4\pi\int d\boldsymbol{r}g_{i}(\boldsymbol{r}-\boldsymbol{r}_{i},R_{0})g_{j}(\boldsymbol{r}-\boldsymbol{r}_{j},R_{0}).
\end{align}
Assuming Gaussian mass density profiles (following e.g. \cite{ghirardi1990continuous, donadi2021underground} $g_{i}(\boldsymbol{r}-\boldsymbol{r}_{i},R_{0})=(2\pi R_{0}^{2})^{-3/2}e^{-\frac{(\boldsymbol{r}-\boldsymbol{r_{i}})^{2}}{2R_{0}^{2}}}$)
and specifying the rate for the mean radii of the atomic orbits we obtain (see Appendix \ref{ratedpc2}): 
\begin{widetext}
\begin{align}\label{ratecdp}
\left. \frac{d\Gamma}{dE} \right|_t^{DP} &= \frac{Ge^2}{12 \pi^{5/2} \epsilon_0 c^3 R_0^3 E} \left\{ N_p^2 + N_e +2 \sum_{o \, o' \, \textrm{pairs}} N_{o} \, N_{o'} \, \frac{\textrm{sin} \left[ \frac{ \beta |\rho_o-\rho_{o'}| \, E}{\hbar \, c} \right]}{\left[ \frac{ \beta |\rho_o-\rho_{o'}| \, E}{\hbar \, c} \right]} e^{-\frac{\beta^2 (\rho_o-\rho_{o'})^2}{4R_0^2}} 
+\right.
\nonumber\\
&\left. + \sum_o N_{o} \, (N_{o} -1) \, e^{-\frac{(\alpha \, \rho_o)^2}{4R_0^2}} \cdot \frac{\textrm{sin} \left( \frac{\alpha \, \rho_o \, E}{\hbar \, c} \right)}{\left( \frac{\alpha \, \rho_o \, E}{\hbar \, c} \right)} -2 N_p \sum_o N_{o} \, \frac{\textrm{sin} \left( \frac{\rho_o \, E}{\hbar \, c} \right)}{\left( \frac{\rho_o \, E}{\hbar \, c} \right)} \cdot 
 e^{-\frac{\rho_o^2}{4R_0^2}} \right\} 
\end{align}
\end{widetext}

Again, the simple rate described by equation \eqref{ratewdpnaive} is a good approximation of Eq. \eqref{ratecdp} for energies belonging to the $\gamma$-rays domain.
%
\begin{figure}
\centering
\includegraphics[width=\columnwidth]{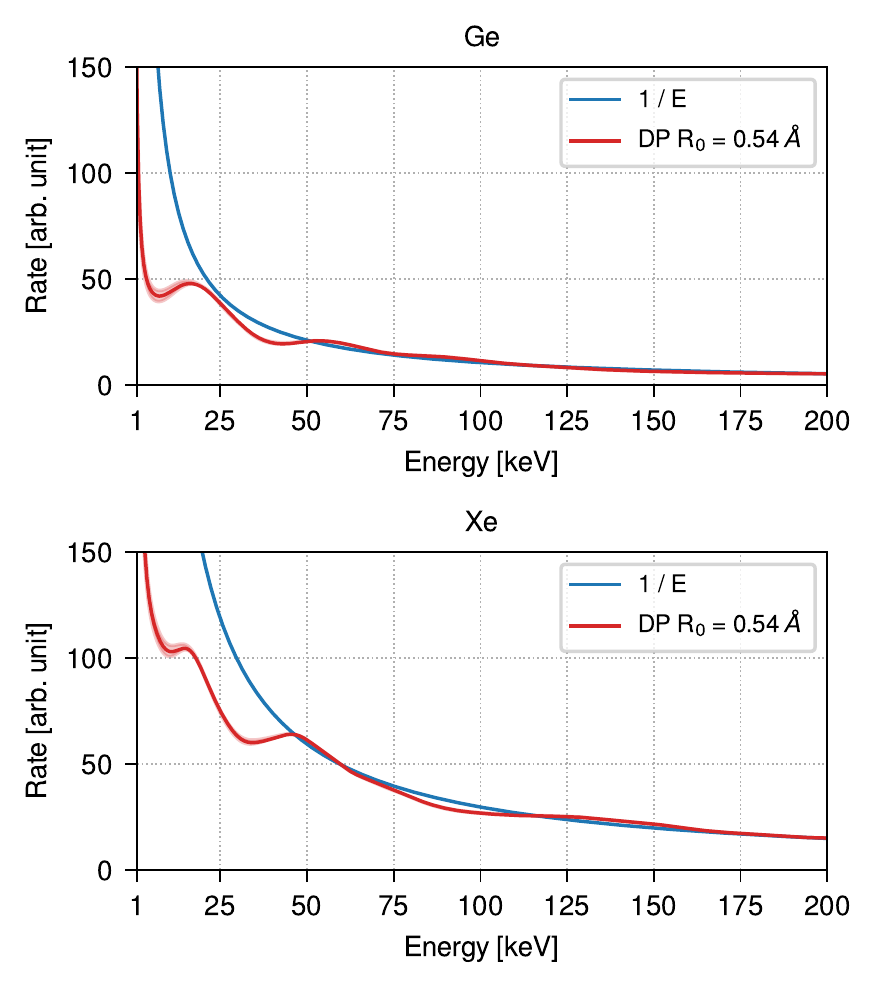}
\caption{Top panel of the Figure shows in blue the $1/E$ dependence Eq. \eqref{ratewdpnaive}, for the spontaneous radiation rate of the DP model, which is only valid in the high-energy domain. This is compared to the general rate Eq. \eqref{ratecdp} (red curve) for a prior value of the correlation length $R_0 = 0.54$ \AA. The distributions are calculated for a Germanium atom and normalized to the common constant pre-factors.
 The bottom panel of the Figure shows the shapes of the same rates, calculated for a Xenon atom.}
\label{comparisondp}
\end{figure}
Figure \ref{comparisondp} compares the general expression of the DP spontaneous emission rate (Eq. \eqref{ratecdp}) given by the red curve, with the simple expression in Eq. \eqref{ratewdpnaive} given by the blue curve. 
The prior value $R_0 = 0.54\;\AA$  
 is chosen, consistently with the result \cite{donadi2021underground}, which is obtained by applying Eq. \eqref{ratewdpnaive} in the $\gamma$-rays regime ($(1-3.8)$ MeV), where Eq. \eqref{ratecsl3} is an excellent approximation.
A strong departure from the approximate expression is evident, at low-energy, for both Germanium (top panel) and Xenon (bottom panel) atoms. A significant dependence of the generalized rate on the atomic number is evident for the DP model as well.\\

{\it Discussion.} --- Figures  
\ref{comparison} and \ref{comparisondp} summarize our findings and unveil the most interesting consequence of the cancellation phenomenon, for which the contribution to the spontaneous radiation emission from oppositely charged particles, whose distance is exceeded by the model's correlation length and by the observed photon wavelength, cancels. 
In the low-energy regime, for correlation lengths of the models of the order, or bigger, than the atomic orbits radii, the shapes predicted for the spontaneous emission rates distributions of the CSL and DP models strongly differ. This is both due to the different mathematical structure of the terms $f_{ij}$ and the different values of the correlation lengths, $r_C$ and $R_0$, of the two models.
In the simple scenario in which $\lambda_\gamma$ is much smaller than the atomic size, any difference is washed-out and the shapes of the spontaneous radiation rates of the two models just differ by a scaling factor. 

An experimental investigation of the spontaneous radiation emitted by the atoms, in the energy range going from few to tens of keV, 
exploiting the outlined phenomenological analysis scheme, and based on the predicted rates Eqs. \eqref{ratecsl5} and \eqref{ratecdp}, 
may be able to disentangle which model describes the data with higher probability. If the measurement is sensitive to the signal of collapse, it should then be also able to recognize the physical mechanism which is responsible for it. Moreover, the experimental sensitivity may be improved, and the systematic effects minimized, by performing a survey on targets of different atomic numbers, exploiting the peculiar impact of the atomic structure on the cancellation phenomenon.    

In this paper we derived  the fundamental theoretical formulas for these future investigations, both for the white noise models and their non-Markovian generalizations.

\begin{acknowledgments}
\noindent
This publication was made possible through the support of Grant 62099 from the John Templeton Foundation. The opinions expressed in this publication are those of the authors and do not necessarily reflect the views of the John Templeton Foundation.
We acknowledge support from the Foundational Questions Institute and Fetzer Franklin Fund, a donor advised fund of Silicon Valley Community Foundation (Grants No. FQXi-RFP-CPW-2008 and FQXi-MGA-2102), and from the H2020 FET TEQ (Grant No. 766900).
We thank: the INFN Institute, for
supporting the research presented in this article and, in particular, the Gran Sasso underground laboratory of INFN, INFN-LNGS, and its Director, Ezio Previtali, the LNGS staff, and the Low Radioactivity laboratory for the experimental activities dedicated to the search for spontaneous radiation.
We thank the Austrian Science Foundation (FWF) which supports the VIP2 project with the grants P25529-N20, project P 30635-N36 and W1252-N27 (doctoral college particles and interactions).
K.P. acknowledges support from the Centro Ricerche Enrico Fermi - Museo Storico della Fisica e Centro Studi e Ricerche ``Enrico Fermi'' (Open Problems in Quantum Mechanics project). S.D. acknowledges support from the Marie
Sklodowska Curie Action through the UK Horizon Europe guarantee administered by UKRI. A.B. acknowledges financial support from the University of Trieste, INFN, the PNRR MUR project PE0000023-NQSTI, and the EIC Pathfinder project QuCoM (GA No. 101046973).
\end{acknowledgments}

\onecolumngrid
\appendix
\section{} \label{fij}
\noindent

We reconsider here the general expression for $f_{ij}$ \cite{donadi2021novel}:   
\begin{equation}
f_{ij} = \sum_{k = x,y,z} \int d^3s \int d^3 s' \, e^{-\frac{(\mathbf{r}_i-\mathbf{r}_j+\mathbf{s}'-\mathbf{s})^2}{4r_C^2}} \,
\frac{\partial \mu_i (\mathbf{s})}{\partial s^k} \frac{\partial \mu_j (\mathbf{s}')}{\partial s'^k},
\end{equation}
$\mu_i$ denoting the mass density of the $i$-th particle, but we don't assume $r_C \gg |\mathbf{r}_i-\mathbf{r}_j|$, condition which was experimentally tested only in the case of a Markovian CSL. Hence
\begin{equation}
f_{ij} = \int d^3s \int d^3 s' \, 
 \mu_i (\mathbf{s}) \mu_j (\mathbf{s}') 
 \left\{ e^{-\frac{(\mathbf{r}_i-\mathbf{r}_j+\mathbf{s}'-\mathbf{s})^2}{4r_C^2}} \frac{1}{2r_C^2} \left[ 3- \frac{(\mathbf{r}_i-\mathbf{r}_j+\mathbf{s}'-\mathbf{s})^2}{2r_C^2} \right] \right\}.
\end{equation}
which reduces to Eq. \eqref{fijcsl2} for point-like mass densities $\mu_i (\mathbf{r}) = m_i \, \delta(\mathbf{r})$.

\section{}\label{whiterate}
\noindent 
By substituting the general expression for $f_{ij}$ given by Eq. \eqref{fijcsl2} in Eq. \eqref{ratecsl1},  and considering the separate contributions of the protons ($i_p,j_p$), of the electrons ($i_e,j_e$) and of the combined electrons-protons emission ($i_p,j_e$) and ($i_e,j_p$) we have:   

\begin{equation}\nonumber
\left. \frac{d\Gamma}{dE} \right|_t^{CSL} =  \frac{\hbar \, \lambda}{12\, \pi^2 \, \epsilon_0 \, c^3 \, m_0^2\, r_C^2 \, E} \cdot  
\end{equation}

\begin{equation}\nonumber
 \left[ \sum_{ip,jp} q_{ip} \, q_{jp} \cdot  \frac{\sin(2\pi |\mathbf{r}_{ip}-\mathbf{r}_{jp}| / \lambda_\gamma)}{2\pi |\mathbf{r}_i-\mathbf{r}_j| / \lambda_\gamma} \, e^{-\frac{(\mathbf{r}_{ip}-\mathbf{r}_{jp})^2}{4r_C^2}} \left( 3 - \frac{(\mathbf{r}_{ip}-\mathbf{r}_{jp})^2}{2r_C^2} \right) \, +
\right. 
\end{equation}

\begin{equation}\nonumber
+ \sum_{ip,je} q_{ip} \, q_{je} \cdot \frac{\sin(2\pi |\mathbf{r}_{ip}-\mathbf{r}_{je}| / \lambda_\gamma)}{2\pi |\mathbf{r}_{ip}-\mathbf{r}_{je}| / \lambda_\gamma} \, e^{-\frac{(\mathbf{r}_{ip}-\mathbf{r}_{je})^2}{4r_C^2}} \left( 3 - \frac{(\mathbf{r}_{ip}-\mathbf{r}_{je})^2}{2r_C^2} \right) \, + 
\end{equation}

\begin{equation}\nonumber
+ \, \sum_{ie,jp} q_{ie} \, q_{jp} \cdot \frac{\sin(2\pi |\mathbf{r}_{ie}-\mathbf{r}_{jp}| / \lambda_\gamma)}{2\pi |\mathbf{r}_{ie}-\mathbf{r}_{jp}| / \lambda_\gamma} \, e^{-\frac{(\mathbf{r}_{ie}-\mathbf{r}_{jp})^2}{4r_C^2}} \left( 3 - \frac{(\mathbf{r}_{ie}-\mathbf{r}_{jp})^2}{2r_C^2} \right) \, +
\end{equation}

\begin{equation}\label{ratecsl4}
+ \, \left.  
\sum_{ie,je} q_{ie} \, q_{je} \cdot \frac{\sin(2\pi |\mathbf{r}_{ie}-\mathbf{r}_{je}| / \lambda_\gamma)}{2\pi |\mathbf{r}_{ie}-\mathbf{r}_{je}| / \lambda_\gamma} \,  e^{-\frac{(\mathbf{r}_{ie}-\mathbf{r}_{je})^2}{4r_C^2}} \left( 3 - \frac{(\mathbf{r}_{ie}-\mathbf{r}_{je})^2}{2r_C^2} \right) \right]
\end{equation}
Let us analyze the three contributions separately:

\begin{enumerate}

    \item The first sum in Eq. \eqref{ratecsl4} is performed on the protons in the nucleus. In the regime that we are analyzing we have $|\mathbf{r}_{ip}-\mathbf{r}_{jp}| / \lambda_\gamma \ll 1$ and $(\mathbf{r}_{ip}-\mathbf{r}_{jp})^2/r_C^2 \ll 1$ hence the sum reduces to:
\begin{equation}\label{sum1}
3 \sum_{ip,jp} q_{ip} \, q_{jp} = 3\,\left(  \sum_{ip} q_{ip} \right)^2 = 3 \, e^2\,N_p^2.
\end{equation}

\item The arguments of the second and third sums are symmetric under the exchange $i \rightarrow j$. For any electron $j_e$ belonging to the $o$-th orbit of the atom we approximate:
\begin{equation}
 |\mathbf{r}_{ip}-\mathbf{r}_{je}| \sim \rho_o  \,\, , \,\, \forall \, ip \,,
\end{equation}
$\rho_o$ representing the mean radius of the orbit hosting the $je$-th electron. Hence, in terms of the energy of the spontaneously emitted photon, and indicating with $N_o$ the number of electrons in the $o$-th orbit, the second and third sum yield:
\begin{equation}\label{sum2}
 -2\, e^2 \, N_p \, \sum_o N_o \, e^{-\frac{\rho_o^2}{4r_C^2}} \, \left( 3 - \frac{\rho_o^2}{2r_C^2} \right) \, \frac{\textrm{sin} \left( \frac{\rho_o \, E}{\hbar \, c} \right)}{\left( \frac{\rho_o \, E}{\hbar \, c} \right)}. 
\end{equation}

\item Concerning the last sum, if $ie=je$ then
\begin{equation}\label{sum3}
3 \sum_{ie,je \, , \, ie=je} q_{ie} q_{je} = 3\,\sum_{ie} q_{ie}^2 = 3\, e^2 \, N_e.
\end{equation}
For $i_e,j_e$ belonging to the same orbit $o$, with $ie\neq je$ we parametrize their distance as $|\mathbf{r}_{ie}-\mathbf{r}_{je}| = \alpha \rho_o$. 
Considered that the total number of pairs in the $o$-th orbit is $N_o\,(N_o-1)$ we find:
%
\begin{equation}\label{sum4}
 e^2 \, \sum_o N_{o} \, (N_{o} -1) \, e^{-\frac{(\alpha \, \rho_o)^2}{4r_C^2}} \cdot \frac{\textrm{sin} \left( \frac{\alpha \, \rho_o \, E}{\hbar \, c} \right)}{\left( \frac{\alpha \, \rho_o \, E}{\hbar \, c} \right)} \cdot  \left(  3 - \frac{( \alpha \, \rho_o)^2}{2 r_C^2}  \right).
\end{equation}
For $i_e,j_e$ belonging to different orbits $o$ and $o'$ we adopt the parametrization  $|\mathbf{r}_{ie}-\mathbf{r}_{je}| = \beta |\rho_o - \rho_{o'}|$. The corresponding contribution to the sum is:
%
%
\begin{equation}\label{sum5}
2 e^2 \, \sum_{o \, o' \, \textrm{pairs}} N_{o} \, N_{o'} \, \frac{\textrm{sin} \left[ \frac{ \beta |\rho_o-\rho_{o'}| \, E}{\hbar \, c} \right]}{\left[ \frac{ \beta |\rho_o-\rho_{o'}| \, E}{\hbar \, c} \right]} e^{-\frac{\beta^2 (\rho_o-\rho_{o'})^2}{4r_C^2}} \, \left(  3 - \frac{ \beta^2 (\rho_o-\rho_{o'})^2}{2r_C^2}  \right).
\end{equation}

\end{enumerate}

By substituting Eqs. \eqref{sum1}-\eqref{sum5} in Eq. \eqref{ratecsl4} Eq. \eqref{ratecsl5} is obtained.
%
Concerning $\alpha$ and $\beta$ their values can be, in principle, estimated from the average distance between two electrons in an atom,
extracted from the position intracule function \cite{gill2003}. This function provides the probability distribution of finding two electrons at a given distance. 
Intracule function calculations are available in the literature for helium, lithium, and beryllium.
In helium, the average distance between two electrons in the 1s shell is known at different levels of approximation for the correlation among electrons \cite{coulson}. Using the correlated value for the inter-electronic distance, $\alpha = 1.47$ for the 1s shell.
From calculations in lithium and beryllium \cite{gill2003} both $\alpha$ and $\beta$ can be obtained. For lithium it is found $\alpha = 1.03$ for the 1s shell and $\beta = 1.04$ for the 1s and 2s shells distance. In the case of beryllium $\alpha = 1.20$ for the 1s shell, $\alpha = 1.11$ for the 2s shell, and $\beta = 1.04$ for the 1s and 2s shells distance.
According to the calculated values, the spontaneous emission rate is shown as a red shaded area in Figure \ref{comparison} for $\beta = 1.04$ and $\alpha$ spanning in the range (1 - 1.5), the red line corresponds to $\alpha = 1.25$.
The implementation of specific intracule functions  for Germanium and Xenon, integrated into a full quantum calculation of the spontaneous radiation rate, is currently under development. 

\section{}\label{ratedpc}
\noindent 

By analogy with the proof outlined in \cite{donadi2021novel} the starting point of the semi-classical calculation of the rate emitted by the charged particles of the atomic system, for a coloured DP model, is the classical expression of the total emitted power:
\begin{equation}
P(t) = R_{sp}^2 \int d\Omega S(R_{sp}^2 \hat{n},t)
\end{equation}
with $S$ being the Poynting vector at time $t$, and the integration being performed on a spherical surface of radius $R_{sp}$. This turns out to be
\begin{equation}
P(t) = \frac{1}{64\,\pi^4\, \epsilon_0 \, c^3} \int_{-\infty}^{+\infty} d\omega \int_{-\infty}^{+\infty} d\nu \, e^{i(\omega+\nu)(t-R_{sp}/c)} \, \sum_{i,j} q_i q_j  \, \mathbb{E}[J_{ij}(\omega,\nu)]  
\end{equation}
with
\begin{equation}\label{Jij}
J_{ij}(\omega,\nu)  = 4 \pi \, \ddot{\mathbf{r}}_i(\omega) \cdot \ddot{\mathbf{r}}_j(\nu) \frac{(b_{ij}^2-1) \, \mathrm{sin}(b_{ij}) + b_{ij} \, \mathrm{cos}(b_{ij})}{b_{ij}^3}
-
4 \pi \, \ddot{{r}}_i^z(\omega) \ddot{{r}}_j^z(\nu) \frac{(b_{ij}^2-3) \, \mathrm{sin}(b_{ij}) + 3b \, \mathrm{cos}(b_{ij})}{b_{ij}^3},
\end{equation}
where $b_{ij}=|\omega \mathbf{r}_i + \nu \mathbf{r}_j|/c$.

In the case of the DP model, the acceleration induced by the collapse is:
\begin{equation}\label{acc}
    \ddot{\boldsymbol{r}}_{i}(t)=-\frac{1}{m_{i}}\int d\boldsymbol{r}\,\left[\nabla_{\boldsymbol{r}_{i}(t)}\mu_{i}(\boldsymbol{r}-\boldsymbol{r}_{i}(t))\right]\phi(\boldsymbol{r},t)
\end{equation}
where 
\begin{equation}
  \mu_{i}(\boldsymbol{r})=m_{i}g_{i}(\boldsymbol{r},R_{0}),  
\end{equation}
with $g_i$ a function describing the mass density of size $R_0$ of the $i$-th particle (which will be specified below, see Eq. (\ref{gdp})) and $\phi(\boldsymbol{r},t)$ a Gaussian noise field with zero average and correlation:
\begin{equation}\label{corrDPw}
\mathbb{E}\left[\phi(\boldsymbol{r},t)\phi(\boldsymbol{r}',t')\right]= G\hbar\frac{\delta(t-t')}{|\boldsymbol{r}-\boldsymbol{r}'|}.
\end{equation}
A delta correlation in time implies a Markovian  dynamics. Similarly to the analysis done for the CSL model, non-Markovianity can be introduced by replacing this correlation function with one describing e.g. an exponentially decaying correlation in time, characterized by a cutoff frequency  $\Omega$, i.e.:
\begin{equation}\label{corrDPnw}
\mathbb{E}\left[\phi(\boldsymbol{r},t)\phi(\boldsymbol{r}',t')\right]=\frac{G\hbar}{|\boldsymbol{r}-\boldsymbol{r}'|}\frac{\Omega}{2}e^{-\Omega|t-t'|}.
\end{equation}

The Fourier transform of the $k$-th component of the acceleration takes the form:
\begin{align}
\ddot{r}_{i}^{k}(\omega)&=-\int dte^{-i\omega t}\int d\boldsymbol{r}\,\left[\frac{\partial}{\partial r_{i}^{k}}g_{i}(\boldsymbol{r}-\boldsymbol{r}_{i},R_{0})\right]\phi(\boldsymbol{r},t)\nonumber
\\
&=\int d\boldsymbol{s}\,\left[\frac{\partial}{\partial s^{k}}g_{i}(\boldsymbol{s},R_{0})\right]\tilde{\phi}(\boldsymbol{s}+\boldsymbol{r}_{i},\omega)
\end{align}
where in the second line we performed the change of variables $\mathbf{s} = \mathbf{r}-\mathbf{r}_i$ and introduced $\tilde{\phi}(\boldsymbol{r},\omega):=\int dte^{-i\omega t}\phi(\boldsymbol{r},t)$, with zero average and correlation
\begin{equation}\nonumber
\mathbb{E}\left[\tilde{\phi}(\boldsymbol{r},\omega)\tilde{\phi}(\boldsymbol{r}',\nu)\right]=2\pi G\hbar\left(\frac{\Omega^{2}}{\omega^{2}+\Omega^{2}}\right)\frac{\delta(\nu+\omega)}{|\boldsymbol{r}-\boldsymbol{r}'|}.
\end{equation}

We can now focus on the average of the scalar product of two accelerations, necessary for computing $J_{ij}(\omega,\nu)$ in Eq. (\ref{Jij}):
\begin{equation}
\mathbb{E}\left[ \ddot{\mathbf{r}}_i(\omega)\cdot
\ddot{\mathbf{r}}_j(\nu)\right] = 2\pi  \hbar G  \delta(\omega+\nu) \frac{\Omega^2}{\Omega^2 + \omega^2} f_{ij} 
\end{equation}
where $f_{ij} = \sum_{k = x,y,z} f_{ij}^k$ with
\begin{equation}\label{fijk}
f_{ij}^k=\int d^3s \int d^3 s' \frac{1}{|\mathbf{r}_i-\mathbf{r}_j+\mathbf{s}-\mathbf{s}'|}
\frac{\partial g_i (\mathbf{s},R_0)}{\partial s^k} \frac{\partial g_j (\mathbf{s}',R_0)}{\partial s'^k}. 
\end{equation}
By assuming $f_{ij}^z = f_{ij}/3$, which holds for spherical symmetric mass distributions, we have:

\begin{equation}
\mathbb{E} \left[ J_{ij} (\omega,\nu)\right] = 8\pi^2 \hbar G  \delta(\omega+\nu) \frac{\Omega^2}{\Omega^2 + \omega^2} f_{ij} \frac{2}{3} \frac{\textrm{sin}(b_{ij})}{b_{ij}}
\end{equation}
and 
\begin{equation}
 P(t) = \frac{G\hbar}{12 \pi^2 \epsilon_0 c^3} \int_{-\infty}^{+\infty} d\omega \, \, \sum_{i,j} q_i\, q_j  f_{ij} \frac{\textrm{sin}(b_{ij})}{b_{ij}} \frac{\Omega^2}{\Omega^2 + \omega^2}.
\end{equation}
By equating the last result for the power to the following expression
\begin{equation}
P(t) =  \int_{0}^{+\infty} d\omega \, \hbar \, \omega \frac{d\Gamma_t}{d\omega},
\end{equation}
the spontaneous emission rate in Eq. \eqref{rate} is obtained.

\section{}\label{ratedpc2}
\noindent

By substituting $\mathbf{s} = \mathbf{r} - \mathbf{r}_i$ and $\mathbf{s}' = \mathbf{r}' - \mathbf{r}_j$ in Eq. \eqref{fijk} and summing over $k=x,y,z$ we have

\begin{align}\label{fij1}
    f_{ij}&=\sum_{k=x,y,z}\int d\boldsymbol{r}\int d\boldsymbol{r}'\left[\frac{\partial}{\partial r^{k}}g_{i}(\boldsymbol{r}-\boldsymbol{r}_{i},R_{0})\right]\left[\frac{\partial}{\partial r'^{k}}g_{j}(\boldsymbol{r}'-\boldsymbol{r}_{j},R_{0})\right]\frac{1}{|\boldsymbol{r}-\boldsymbol{r}'|}=\nonumber
    \\
    &=\int d\boldsymbol{r}\int d\boldsymbol{r}'g_{i}(\boldsymbol{r}-\boldsymbol{r}_{i},R_{0})g_{j}(\boldsymbol{r}'-\boldsymbol{r}_{j},R_{0})\sum_{k=x,y,z}\left[\frac{\partial}{\partial r'^{k}}\frac{\partial}{\partial r^{k}}\frac{1}{|\boldsymbol{r}-\boldsymbol{r}'|}\right]=\nonumber
    \\
    &=\int d\boldsymbol{r}\int d\boldsymbol{r}'g_{i}(\boldsymbol{r}-\boldsymbol{r}_{i},R_{0})g_{j}(\boldsymbol{r}'-\boldsymbol{r}_{j},R_{0})\left[-\nabla_{r}^{2}\frac{1}{|\boldsymbol{r}-\boldsymbol{r}'|}\right]=\nonumber\\
    &=\int d\boldsymbol{r}\int d\boldsymbol{r}'g_{i}(\boldsymbol{r}-\boldsymbol{r}_{i},R_{0})g_{j}(\boldsymbol{r}'-\boldsymbol{r}_{j},R_{0})\left[4\pi\delta(\boldsymbol{r}-\boldsymbol{r}')\right]=\nonumber
    \\
    &=4\pi\int d\boldsymbol{r}g_{i}(\boldsymbol{r}-\boldsymbol{r}_{i},R_{0})g_{j}(\boldsymbol{r}-\boldsymbol{r}_{j},R_{0}).
\end{align}
For Gaussian mass density profiles:

\begin{equation}\label{gdp}
g_{i}(\boldsymbol{r}-\boldsymbol{r}_{i},R_{0})=\frac{1}{(2\pi R_{0}^{2})^{3/2}}e^{-\frac{(\boldsymbol{r}-\boldsymbol{r_{i}})^{2}}{2R_{0}^{2}}},
\end{equation}
simple steps lead to 
\begin{equation}\label{fij2}
 f_{ij} = \frac{1}{2\pi^{1/2} R_0^3} e^{-\frac{(\mathbf{r}_i-\mathbf{r}_j)^2}{4R_0^2}}.
\end{equation}
By substituting Eq. \eqref{fij2} in Eq. \eqref{rate} we get: 
\begin{equation}
\left. \frac{d\Gamma}{dE} \right|_t^{DP} = \frac{G}{12 \pi^{5/2} \epsilon_0 c^3 R_0^3 E}  \sum_{i,j} q_i\, q_j \,  e^{-\frac{(\mathbf{r}_i-\mathbf{r}_j)^2}{4R_0^2}}  \, \frac{\textrm{sin}\left( \frac{|\mathbf{r}_i-\mathbf{r}_j| \, E}{\hbar \, c} \right)}{\left( \frac{|\mathbf{r}_i-\mathbf{r}_j| \, E}{\hbar \, c} \right)}.
\end{equation}
By repeating the steps (1)-(3) which lead from Eq. \eqref{ratecsl4} to \eqref{ratecsl5}, Eq. \eqref{ratecdp} is obtained.

\bibliographystyle{apsrev4-2}
\bibliography{ref}
\end{document}